\documentclass[aps,prl,twocolumn,showpacs,superscriptaddress]{revtex4-1}
\usepackage{amsmath,amssymb}
\usepackage{graphicx}
\usepackage{dcolumn}
\usepackage{bm}

\begin{document}
\title{Comment on ``Exact Classification of Landau-Majorana-St\"uckelberg-Zener Resonances by Floquet Determinants''}

\author{Qing-Wei Wang}
\email[]{qwwang@theory.issp.ac.cn}
\affiliation{Department of Physics, Renmin University of China, Beijing 100872, China}
\affiliation{Key Laboratory of Materials Physics, Institute of Solid State Physics, Chinese Academy of Sciences, Hefei 230031, P. R. China}
\author{Yu-Liang Liu}
\affiliation{Department of Physics, Renmin University of China, Beijing 100872, China}

\date{\today}
%
%
%

\maketitle
In a recent Letter \cite{PRL111.130405}, Ganeshan \emph{et al.} present a general framework to classify the resonance structure of Landau-Majorana-St\"uckelberg-Zener interferometry into three basic categories distinguished by whether these resonances correspond to periodic or nonperiodic quantum evolution. In this Comment, we show that their identification of the real resonances in the regime of small drive amplitude is incorrect.

\emph{Analytical argument}.---We consider the example of monochromatic driving, $J(t)=\epsilon+A\sin(\omega t +\varphi)$, and the special case with $\varphi=-\pi/2$. We will prove that the real resonances only occur for $2\epsilon/\omega=$ integer, no matter $A/\omega$ is large or small. According to Floquet analysis, the two Floquet solutions of the Schr\"odinger equation admit Fourier series expansions, and the corresponding Fourier coefficients satisfy a difference equation given by Eq. (9) in Ref. \cite{PRL111.130405}. Standard asymptotic analysis on this difference equation \cite{Elaydi2005} tells us that there are two linearly independent solutions in the $n\rightarrow \infty$ limit, with the asymptotic behavior
\begin{equation}
  a^{(+1)}_{n+1}/a^{(+1)}_{n} \sim A/n, \quad a^{(+2)}_{n+1}/a^{(+2)}_{n} \sim n/A.
\end{equation}
Obviously $a^{(+1)}$ is a \emph{minimal solution}. Similarly, in the $n\rightarrow -\infty$ limit, we have another two linearly independent solutions with the asymptotic behavior
\begin{equation}
  a^{(-1)}_{n-1}/a^{(-1)}_{n} \sim A/n, \quad a^{(-2)}_{n-1}/a^{(-2)}_{n} \sim n/A,
\end{equation}
where $a^{(-1)}$ is a minimal solution.

When $k/2+\epsilon/\omega \neq$ integer, all the subdiagonal elements of $M_k$ [Eq.(1) in the Supplementary Material of Ref.\cite{PRL111.130405}] do not vanish, so any two adjacent elements $(a_n,a_{n+1})$ can determine the whole vector $a$ completely. This means that the two-dimensional solution space $\{a^{(+1)}, a^{(+2)}\}$ is equivalent to the space $\{a^{(-1)}, a^{(-2)}\}$. The \emph{physical solution} should be minimal on the $n\rightarrow \pm\infty$ directions, so if such a solution exist, it must be \emph{unique}. Then no real resonance can occur since at real resonance the quasienergy level should be twofold degenerate.

When $k/2+\epsilon/\omega =$ integer, for example, $k/2+\epsilon/\omega=0$, the 0-th column of $M_k$ would have only one nonvanishing element. The set of equations $M_k a=0$ would decouple into two independent sets of equations: one for the semi-infinite vector $(a_0, a_1, \cdots)$ and the other for $(\cdots, a_{-1}, a_0)$.  Real resonance can occur when these two sets of equations have minimal solutions simultaneously.

\begin{figure}
  \centering
  \includegraphics[width=0.48\textwidth]{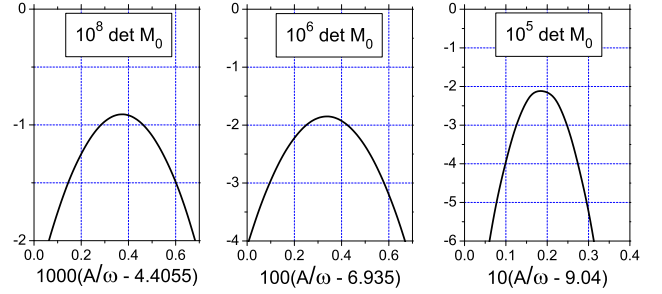}\\
  \caption{The Floquet determinant $\det M_0$ near the three ``real resonances'' marked in Fig.3 of Ref.\cite{PRL111.130405} with $\epsilon/\omega=\sqrt{2}$ and $h/\omega=5$. The precisions are respectively $10^{-10}, 10^{-8}$ and $10^{-6}$.} \label{fig:detM0}
\end{figure}

\emph{Numerical argument}.---According to Ref.\cite{PRL111.130405}, the real resonances can be determined from the roots of the Floquet determinant, i.e., $\det M_k=0$ [Eq.(10) in Ref.\cite{PRL111.130405}], with $k=0$ or $1$. However, this determinant does not exist; a proper normalization should be done. Here we take the normalized form of $M_k$ as $[C_{m,n}]_{-\infty}^\infty$, an infinite matrix formed of the numbers $C_{m,n}$, with $C_{0,\pm1}=(A/\omega)(k/2+\epsilon/\omega\mp1), C_{0,0}=k^2/4-(\epsilon^2+h^2)/\omega^2$, and
\begin{eqnarray*}
  C_{m,m\pm1} &=& (A/\omega) \left(k/2+\epsilon/\omega-m \mp1 \right)/m^2,  \\
  C_{m,m} &=& \left[(m-k/2)^2-\left(\epsilon^2+h^2 \right)/\omega^2 \right]/m^2,
\end{eqnarray*}
for $m\neq 0$, and all other matrix elements vanish. Using an algorithm given in the Supplementary Material \cite{*[{See Supplemental Material at [URL] for an algorithm designed to compute the determinant of a tridiagonal infinite matrix}][{}] sm} we compute the determinant $\det M_0$ near the three ``real resonances'' marked in Fig.3 of Ref.\cite{PRL111.130405} and plot the results in Fig.\ref{fig:detM0}. Our high-precision computations show that there are no real roots of $\det M_0$ near these points, therefore the corresponding  resonances should be not real but complex. Similarly, the real resonances marked in any figure with $2\epsilon/\omega\neq$ integer should be complex resonances.

In conclusion, we have shown that the identification of the real resonances based on Floquet determinants in Ref.\cite{PRL111.130405} is incorrect in the regime of small drive amplitude. The real resonances should only occur for $2\epsilon/\omega=$ integer.



\begin{thebibliography}{0}%
\makeatletter
\providecommand \@ifxundefined [1]{%
 \@ifx{#1\undefined}
}%
\providecommand \@ifnum [1]{%
 \ifnum #1\expandafter \@firstoftwo
 \else \expandafter \@secondoftwo
 \fi
}%
\providecommand \@ifx [1]{%
 \ifx #1\expandafter \@firstoftwo
 \else \expandafter \@secondoftwo
 \fi
}%
\providecommand \natexlab [1]{#1}%
\providecommand \enquote  [1]{``#1''}%
\providecommand \bibnamefont  [1]{#1}%
\providecommand \bibfnamefont [1]{#1}%
\providecommand \citenamefont [1]{#1}%
\providecommand \href@noop [0]{\@secondoftwo}%
\providecommand \href [0]{\begingroup \@sanitize@url \@href}%
\providecommand \@href[1]{\@@startlink{#1}\@@href}%
\providecommand \@@href[1]{\endgroup#1\@@endlink}%
\providecommand \@sanitize@url [0]{\catcode `\\12\catcode `\$12\catcode
  `\&12\catcode `\#12\catcode `\^12\catcode `\_12\catcode `\%12\relax}%
\providecommand \@@startlink[1]{}%
\providecommand \@@endlink[0]{}%
\providecommand \url  [0]{\begingroup\@sanitize@url \@url }%
\providecommand \@url [1]{\endgroup\@href {#1}{\urlprefix }}%
\providecommand \urlprefix  [0]{URL }%
\providecommand \Eprint [0]{\href }%
\providecommand \doibase [0]{http://dx.doi.org/}%
\providecommand \selectlanguage [0]{\@gobble}%
\providecommand \bibinfo  [0]{\@secondoftwo}%
\providecommand \bibfield  [0]{\@secondoftwo}%
\providecommand \translation [1]{[#1]}%
\providecommand \BibitemOpen [0]{}%
\providecommand \bibitemStop [0]{}%
\providecommand \bibitemNoStop [0]{.\EOS\space}%
\providecommand \EOS [0]{\spacefactor3000\relax}%
\providecommand \BibitemShut  [1]{\csname bibitem#1\endcsname}%
\let\auto@bib@innerbib\@empty
\end{thebibliography}%


\begin{thebibliography}{3}%
\makeatletter
\providecommand \@ifxundefined [1]{%
 \@ifx{#1\undefined}
}%
\providecommand \@ifnum [1]{%
 \ifnum #1\expandafter \@firstoftwo
 \else \expandafter \@secondoftwo
 \fi
}%
\providecommand \@ifx [1]{%
 \ifx #1\expandafter \@firstoftwo
 \else \expandafter \@secondoftwo
 \fi
}%
\providecommand \natexlab [1]{#1}%
\providecommand \enquote  [1]{``#1''}%
\providecommand \bibnamefont  [1]{#1}%
\providecommand \bibfnamefont [1]{#1}%
\providecommand \citenamefont [1]{#1}%
\providecommand \href@noop [0]{\@secondoftwo}%
\providecommand \href [0]{\begingroup \@sanitize@url \@href}%
\providecommand \@href[1]{\@@startlink{#1}\@@href}%
\providecommand \@@href[1]{\endgroup#1\@@endlink}%
\providecommand \@sanitize@url [0]{\catcode `\\12\catcode `\$12\catcode
  `\&12\catcode `\#12\catcode `\^12\catcode `\_12\catcode `\%12\relax}%
\providecommand \@@startlink[1]{}%
\providecommand \@@endlink[0]{}%
\providecommand \url  [0]{\begingroup\@sanitize@url \@url }%
\providecommand \@url [1]{\endgroup\@href {#1}{\urlprefix }}%
\providecommand \urlprefix  [0]{URL }%
\providecommand \Eprint [0]{\href }%
\providecommand \doibase [0]{http://dx.doi.org/}%
\providecommand \selectlanguage [0]{\@gobble}%
\providecommand \bibinfo  [0]{\@secondoftwo}%
\providecommand \bibfield  [0]{\@secondoftwo}%
\providecommand \translation [1]{[#1]}%
\providecommand \BibitemOpen [0]{}%
\providecommand \bibitemStop [0]{}%
\providecommand \bibitemNoStop [0]{.\EOS\space}%
\providecommand \EOS [0]{\spacefactor3000\relax}%
\providecommand \BibitemShut  [1]{\csname bibitem#1\endcsname}%
\let\auto@bib@innerbib\@empty
\bibitem [{\citenamefont {Ganeshan}\ \emph {et~al.}(2013)\citenamefont
  {Ganeshan}, \citenamefont {Barnes},\ and\ \citenamefont
  {Das~Sarma}}]{PRL111.130405}%
  \BibitemOpen
  \bibfield  {author} {\bibinfo {author} {\bibfnamefont {S.}~\bibnamefont
  {Ganeshan}}, \bibinfo {author} {\bibfnamefont {E.}~\bibnamefont {Barnes}}, \
  and\ \bibinfo {author} {\bibfnamefont {S.}~\bibnamefont {Das~Sarma}},\ }\href
  {\doibase 10.1103/PhysRevLett.111.130405} {\bibfield  {journal} {\bibinfo
  {journal} {Phys. Rev. Lett.}\ }\textbf {\bibinfo {volume} {111}},\ \bibinfo
  {pages} {130405} (\bibinfo {year} {2013})}\BibitemShut {NoStop}%
\bibitem [{\citenamefont {Elaydi}(2005)}]{Elaydi2005}%
  \BibitemOpen
  \bibfield  {author} {\bibinfo {author} {\bibfnamefont {S.}~\bibnamefont
  {Elaydi}},\ }\href@noop {} {\emph {\bibinfo {title} {An Introduction to
  Difference Equations}}}\ (\bibinfo  {publisher} {Springer},\ \bibinfo {year}
  {2005})\BibitemShut {NoStop}%
\bibitem [{sm()}]{sm}%
  \BibitemOpen
  \href@noop {} {}\BibitemShut {NoStop}%
\end{thebibliography}

%

\end{document}